\documentclass[aps,twocolumn,superscriptaddress,showpacs]{revtex4}
\usepackage{graphicx}
\usepackage{amsmath}
\usepackage{color}
\usepackage{enumitem}
\usepackage{epsfig}

\begin{document}
\title{Self-assembly in chains, rings and branches: a single component system with 
two critical points}
\author{Lorenzo Rovigatti} 
\affiliation{ {Dipartimento di Fisica,  {\em Sapienza} Universit\`a di Roma, Piazzale A. Moro 2, 00185 Roma, Italy} }
\author{Jos\'e Maria Tavares}
\affiliation{Instituto Superior de Engenharia de Lisboa - ISEL,
Rua Conselheiro Em\'{\i}dio Navarro 1, P-1950-062 Lisbon, Portugal }
\affiliation{
Centro de F\'{\i}sica Te\'{o}rica e Computacional,
Avenida Professor Gama Pinto 2, P-1649-003 Lisbon, Portugal
}
\author{  Francesco Sciortino} 
\affiliation{ {Dipartimento di Fisica and  CNR-ISC, {\em Sapienza} Universit\`a di Roma, Piazzale A. Moro 2, 00185 Roma, Italy} }

\begin{abstract}
We study 
the interplay between phase separation and self-assembly in chains, rings and branched structures in 
a model of  particles with dissimilar patches.   We extend
  Wertheim's first order perturbation theory to include the effects of ring formation and
 theoretically investigate the thermodynamics of the model.  We find a peculiar shape for the vapor-liquid
 coexistence,
 featuring re-entrant behavior in both phases  
 and two critical points, despite the single-component nature of the system.
The emergence of the lower critical point is 
caused by the self-assembly of rings taking place in the vapor, generating a phase with  lower energy and lower entropy than the liquid. Monte Carlo simulations of the same model  fully support  these unconventional theoretical predictions.  
\end{abstract}



\maketitle
Understanding the competition between self-assembly and phase-separation  is central in today's research in
different fields encompassing biology, soft-matter, material science, statistical mechanics. Self-assembly of finite-size aggregates requires strong interaction energies compared to the thermal energy to guarantee that the generated structure is persistent.  As a result, self-assembly competes with the ubiquitous macroscopic phase separation, the low-temperature 
tendency common to atoms, molecules and larger particles to maximize the number of bonded neighbours and minimize the potential energy, giving rise to a condensed (liquid) state~\cite{Hansennew}.  When particles can arrange themselves
into low energy and weakly interacting  finite size aggregates,  self-assembly can completely suppress
phase separation~\cite{panagiotopoulos,likos_telechelic,reinhardt}.  Soft matter and biology offers several examples of  three-dimensional stable aggregates  (e.g. micelles, vesicles, capsids\cite{jones,PhysRevE.68.051910}) which constitute the stable phase in wide temperature and density regions. 

The possibility of hampering the formation of large amorphous aggregates in favor of
ordered structures is often facilitated by  the presence of strong, directional and saturable interactions (limited valence)~\cite{GlotzNatMat04,bianchi2006,likos}. 
The renewed focus on the role of directional interactions,  stimulated by the synthesis of new-generation patchy colloids~\cite{Manoh_03,Paw10a,granick,granicknature}, has deepened our understanding not only of the role of the valence on the gas-liquid phase separation\cite{Zacca1} but also on the competition between self-assembly and phase separation.  Two recent investigations have provided insights particularly relevant for this work: (i) a  numerical study
of Janus colloids in which  a gas-liquid  critical point and a self-assembly process are 
simultaneously observed~\cite{janusprl,pccpjanus}. In this model,  the formation of energetically stable vesicles stabilizes at low temperature $T$ the gas-phase; (ii) a  study of  particles with  dissimilar patches~\cite{molphyslisbon,prl-lisbona,lisbona_lungo}, promoting respectively chaining and branching, specifically designed to reproduce a mean-field model introduced by Safran and Tlustly~\cite{safrannew} to describe the phase behavior of dipolar fluids.  Here
branching produces a gas-liquid critical point, but on cooling 
the formation of  energetically favored chains stabilizes, this time, the liquid-phase. In both models,
at low $T$ self-assembly  opens up a low-density region of thermodynamic stability in which no macroscopic phase separation takes place.  

In this Letter we investigate the competition between phase separation and
self-assembly in a  model of patchy particles with dissimilar patches in three dimensions, specifically designed  to favor self-assembly in energetically stable   ring structures.  Extending   Wertheim's theory to the case in which chains, branched structures  and
rings coexist, we are able to solve the model, providing a parameter-free analytic formulation of a thermodynamic system 
in which phase separation is suppressed at low $T$ by self-assembly.
We indeed find  theoretically, and confirm numerically, that
macroscopic phase separation can be limited to intermediate $T$
 via the intervention of a closed coexistence loop, despite the one-component nature of the model. The  low $T$ region of the phase diagram  is thus devoid of  any thermodynamically unstable  region.  The amplitude of the closed loop in the $T-\rho$ plane
 shrinks progressively on weakening branching or, equivalently, on increasing ring formation, providing a neat
 mechanism for understanding the smooth disappearance of the   phase separation.
  Our results are consistent with and provide a theoretical base to
a recent numerical investigation of a ring-forming model on a lattice in two-dimensions~\cite{almarza_rings}.   
Finally, our results provide a reference system for understanding the low $T$ behavior of dipolar hard spheres (DHS), hard spheres with an embedded central point magnetic dipole, for which recent numerical studies have reported self-assembly into ring structures, possibly suppressing phase separation~\cite{dhs_prl,dhs_soft_matter}.

{\it Model:} We study a modification of the  model of patchy particles with dissimilar patches which was specifically designed to present a competition between  chains and branched structures. In Ref.~\cite{prl-lisbona} particles were modelled as hard spheres 
 of diameter $\sigma$, with patches of two types on their surface: 2 patches of type A  on the poles and $n$ patches of type B equally spaced  over the equator. 
When two patches of type $\alpha$ and $\beta$ ($\alpha,\beta  \in A,B$), are close enough and properly oriented (see 
supplementary information (S.I.) for details), a bond $\alpha\beta$ is formed. Each bond 
$\alpha\beta$ is characterized by a bonding energy  $\epsilon_{\alpha\beta}$ (the decrease in energy upon bond formation) and a bonding volume $v_{\alpha\beta}$.  
By setting $\epsilon_{BB}=0$, only $AA$ and $AB$ bonds are retained: $AA$ bonds correspond to linear 
self assembly (in chains or rings) and $AB$ bonds to branching points or junctions. 
As in the previous study~\cite{prl-lisbona}, we set $\epsilon_{AB}=0.37\epsilon_{AA}$ to make the formation of chains energetically favorable at low $T$ and $n=9$ and $v_{AB}\gg v_{AA}$ to make branching entropically favorable \cite{lisbona_lungo}.  
The positions and sizes of the patches are chosen to satisfy the single-bond-per-patch condition.

Locating the two A patches on the poles  (as in \cite{prl-lisbona}) generates very long and persistent chains, effectively suppressing the formation of rings.  In this case the competition between chains and branching  originates a  
liquid vapor phase separation \cite{prl-lisbona,lisbona_lungo,almarza_tavares} in which the   gas-liquid binodal is re-entrant: the density of the coexisting liquid approaches the density of the coexisting gas. 
In the present work we add the possibility of ring formation by selecting an 
 off pole position of the $A$ patches.  This very simple modification alters the persistence length of the
 chains and favors the formation of rings,  structures entropically unfavored but energetically stabilized by the additional bond  compared to chains of the same size.  Moreover, self assembly in rings  decreases the possibility of forming junctions. In fact, when two chains assemble to form a longer chain, 2 $A$ unbonded patches (capable of forming a junction) still remain; but when two chains join to form a ring, $A$ patches saturate and become unavailable to form junctions. 

{\it Theory:} 
Wertheim's theory has successfully described the effect of association in the phase diagram of molecular fluids 
\cite{jacksonSAFT} and has, more recently, provided free parameter descriptions of the properties of 
patchy particle models~\cite{lisbona_lungo,almarza_tavares,BianchiJCP07,bianchi2006}.
Extensions of the theory to include the effects of ring formation have been limited to the cases of particles with two patches \cite{sear1994,galindo2002,avlund2011,duerings,marshall_wertheim} (i.e. to the competition between chain and ring formation, but no branching) or with one patch, but allowing the formation of rings with 3 particles by double bonding of the patches
\cite{marshall2012}. In any case,  a single energy scale (one type of patch) was considered and no study of phase diagrams was carried out, since the structures formed - rings and chains only - do not phase separate.
As explained in detail in the S.I.  
we extend Wertheim's first order 
perturbation theory for patchy particles models with dissimilar patches to the case where rings are formed. 
Rings can be of any size $i$ and are considered to be sequences of particles bonded through $i$ consecutive 
AA bonds, regardless of the state of $B$ sites. The free energy (per particle) is the sum of the reference hard-sphere free energy and a bonding contribution,
\begin{equation}
\label{fbmain}
\beta f_b=\ln (Y X_B^n)-X_A-\frac{n}{2}X_B+\frac{n}{2}+1-\frac{G_0}{\rho},
\end{equation}
where $X_A$ and $X_B$ are the fraction of unbonded patches of type A and B, respectively,   $Y$ is 
the fraction of particles with  both patches A unbonded and $G_0$ is the number density of rings.
$X_A$, $X_B$ and $Y$ are calculated as a function of the number density 
$\rho\equiv N/V$ and of $T$ using
the laws of mass action,
\begin{equation}
\label{lma1main} 
1-\frac{X_A^2}{Y}=\frac{G_1}{\rho}
\end{equation}
\begin{equation}
\label{lma2main}
\frac{X_A}{Y}-2n\rho\Delta_{AB}X_B-2\rho\Delta_{AA}X_A=1,
\end{equation}
\begin{equation}
\label{lma3main}
2\rho\Delta_{AB}X_AX_B+X_B=1.
\end{equation}
$\Delta_{\alpha\beta}$ are integrals of the Mayer function of the interaction between patches $\alpha$ and $\beta$, weighted by the pair correlation function of the reference system \cite{lisbona_lungo}. $G_0$ 
and $G_1$ are the zero and first moment of 
the density size distribution of rings (they are a function of $Y,\rho$ and $\Delta_{AA}$, see S.I.).
Thus  the fraction of particles that belong to rings $f_{rings}$  is  $ G_1/\rho$. 
This set of equations defines the thermodynamics of the model, within Wertheim's first order perturbation theory extended to include rings; the values of $X_A$, $X_B$ and $G_1/\rho$ provide information about the self-assembled structures.

\begin{figure}
\centerline{\includegraphics[width=3.3in]{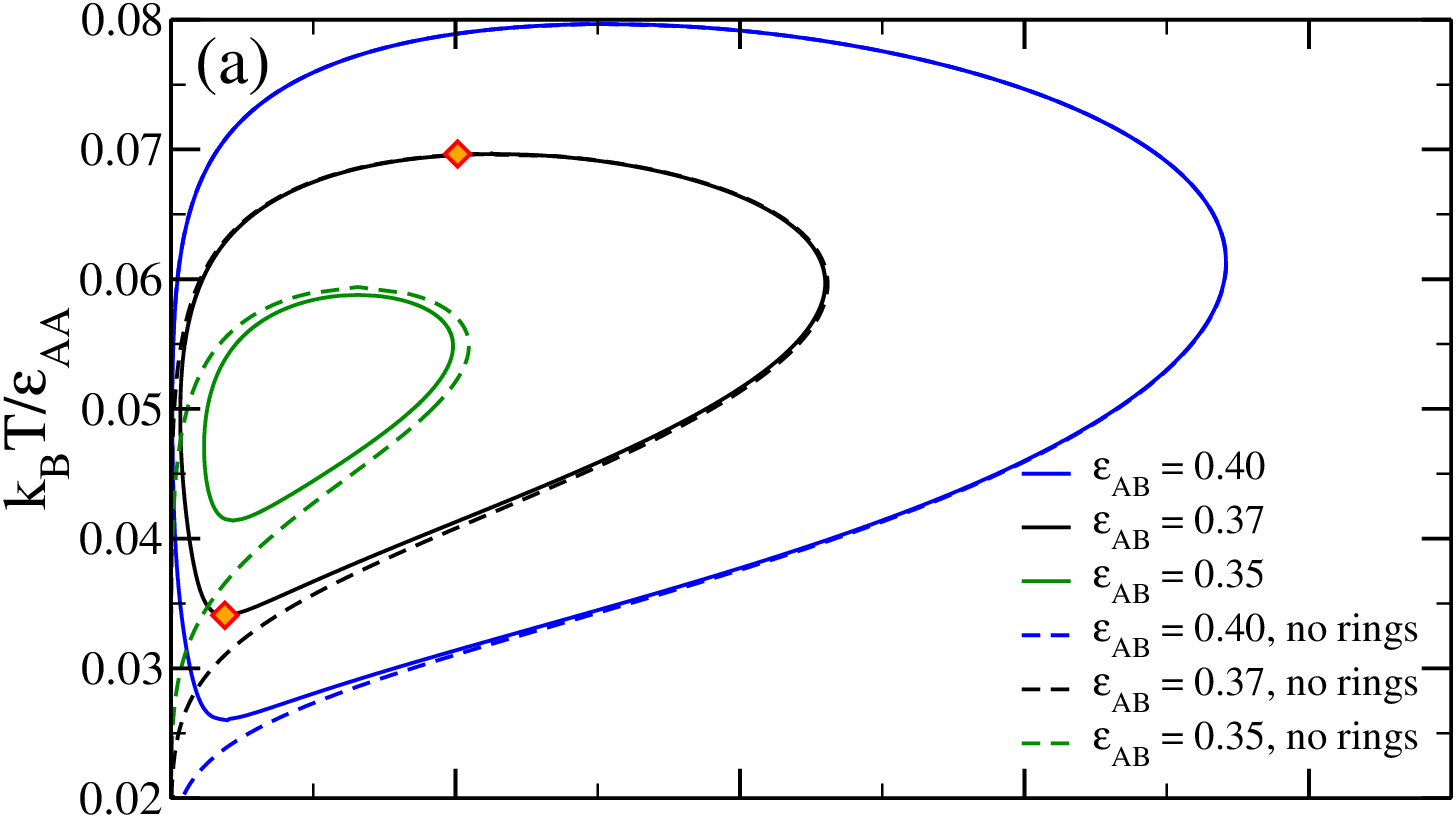}}
~\\[0.01cm]
\centerline{\includegraphics[width=3.3in]{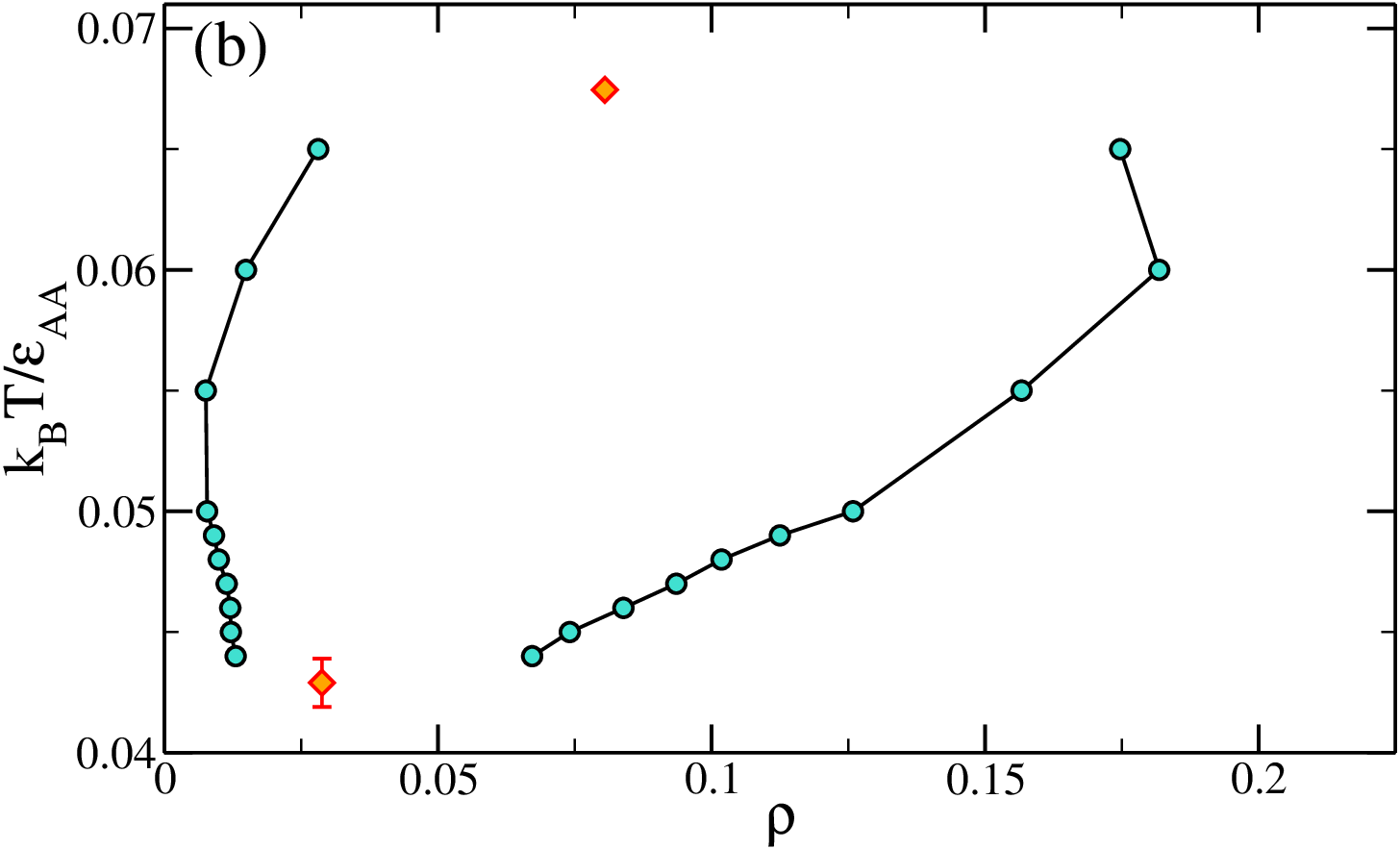}}
\caption{(a) Phase diagram of the model in the $(\rho,T)$ plane, with (full lines) and without (dashed lines) ring clusters for three different values of the branching energy $\epsilon_{AB}$. Lines 
are results from Wertheim's theory. Symbols indicate the position of the upper and lower critical points for the for $\epsilon_{AB}=0.37\epsilon_{AA}$ system. Notice, in the case of rings, the existence of two critical points 
(predicted by theory) and the re-entrant behavior of both the liquid and the vapor parts of the binodal (predicted by theory and confirmed by simulations). (b) Numerically estimated phase diagram of the $\epsilon_{AB}=0.37\epsilon_{AA}$ model. The  critical points are marked by red diamonds. 
}
\label{fig:phd}
\end{figure}

{\it Results:} Fig.~\ref{fig:phd} shows the theoretical results for the gas-liquid coexistence, with and without rings.
When rings are present the theory predicts, in a one-component system, a closed loop and the existence of two critical points: $T_{c,u}$ , the critical temperature of the upper critical point, and $T_{c,l}$ that of the lower critical point.  The coexistence curve is characterized by a re-entrant behavior both in 
 the liquid and in the vapor sides.  Indeed,
when $T$ is decreased below a certain value, the density of the liquid decreases and the density of the vapor increases.
 Coexistence is present only for intermediate $T$.   Below $T< T_{c,l}$ the system remains homogeneous
 for all $T$, completely suppressing any phase separation.   Varying the parameters of the model ($\epsilon_{AB}$ in the figure),
 the closed loop can be progressively shrunk up to the point it disappears, leaving a system for which self-assembly is
 the unique mechanism for aggregation.
  
\begin{figure}[t]
\centerline{\includegraphics[width=3.4in]{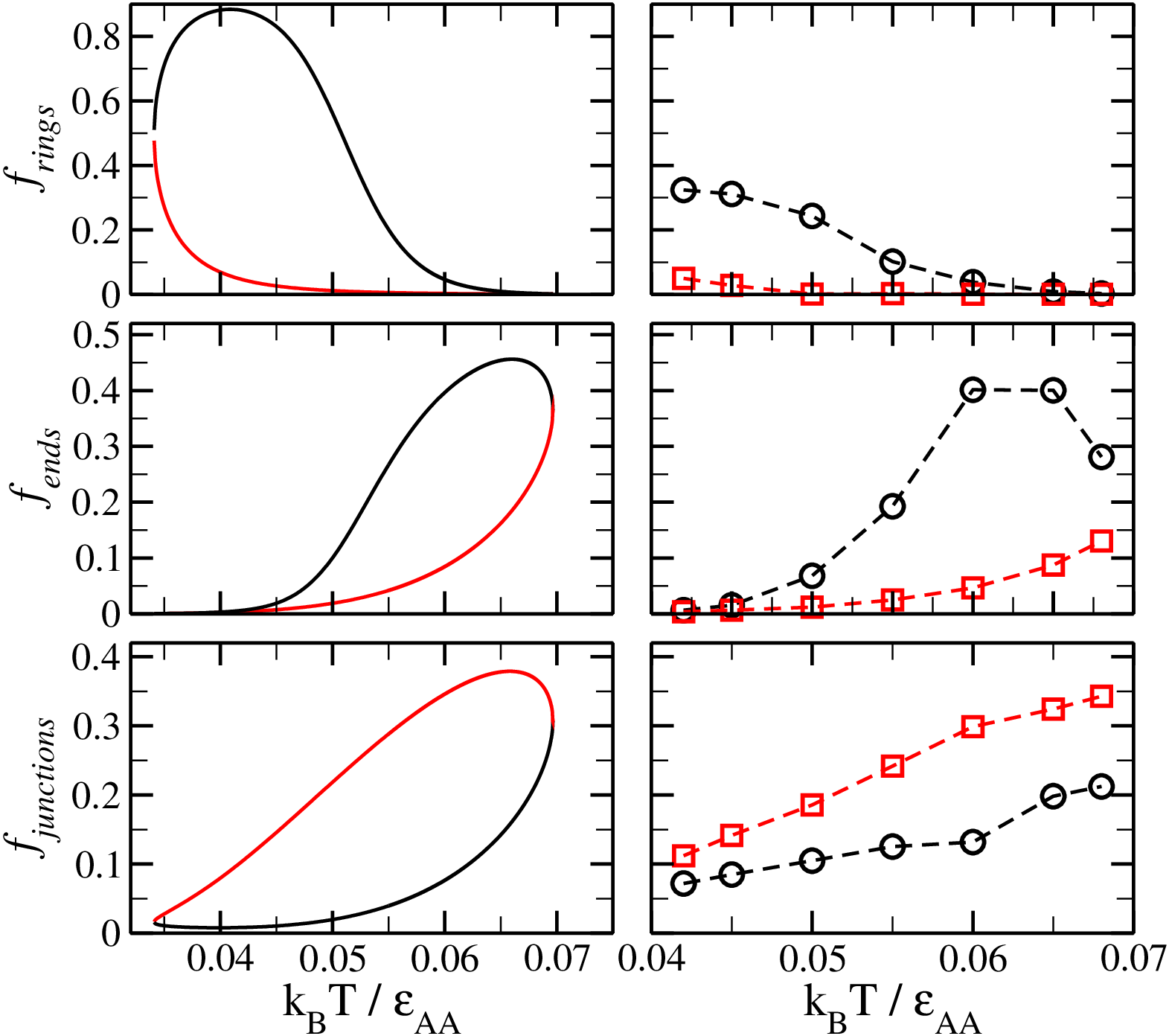}}
\caption{Structural properties calculated at coexistence using Wertheim's theory (left panels, solid lines) and simulation results (right panels, symbols). From top to bottom: fraction of  particles in rings $f_{rings}$, fraction of chain ends $f_{ends}$, and fraction of junctions $f_{junctions}$. Black (red) lines and symbols correspond to the vapor (liquid) phase. }
\label{fig:frnenj}
\end{figure}  
  
The structure of the coexisting phases may be investigated through
the calculation at coexistence of the fraction of particles in rings, $f_{rings}$, the fraction of ends  $f_{ends}$, defined as the
 number of unbonded A patches per particle $f_{ends}\equiv 2X_A$~\cite{lisbona_lungo}, and the fraction of junctions $f_{junctions}$,
 defined as the number of bonded B patches per particle  $f_{junction}\equiv n(1-X_B)$~\cite{lisbona_lungo}.
Fig.~\ref{fig:frnenj} shows the $T$-dependence of $f_{rings}$, $f_{ends}$  and $f_{junctions}$ along the
coexistence curve,  for both gas and liquid.  Close to $T_{c,u}$ both phases exhibit almost no rings: the vapor is characterized
by smaller $f_{junction}$ and larger $f_{ends}$, meaning that it is formed by relatively short and isolated chains, while the liquid contains larger branched chains connected by junctions (a network fluid). 
Coexistence is then obtained between a low density gas of short chains with a few junctions and a network of long chains connected through junctions \cite{safrannew,safran,lisbona_lungo,prl-lisbona}. 
Upon cooling, in the vapor phase $f_{rings}$  increases to significant values, $f_{ends}$ decreases significantly, and $f_{junctions}$ decreases slightly; this means that mostly isolated chains are self-assembling into rings. On the other hand, in the liquid $f_{rings}$ remains small and $f_{ends}$ and $f_{junctions}$ decrease slightly, thus meaning that the network of chains is formed by longer chains and fewer junctions. Therefore, coexistence at these intermediate $T$ is between a vapor where rings dominate, and a network fluid formed by chains and junctions. Finally, close to $T_{c,d}$, as $T$  
decreases, $f_{rings}$ decreases for the vapor phase and increases for the liquid phase, $f_{junctions}$ slightly increases for the vapor phase and decreases for the liquid phase, while $f_{ends}\approx 0$ for both phases. Thus, these phases are evolving to the formation of a network of fully connected chains and rings 
(i.e. with practically all patches A bonded, since $f_{ends}\approx 0$), with more particles in rings (chains) in the vapor (liquid) phase.   

\begin{figure}[h!]
\centerline{\includegraphics[width=3.3in]{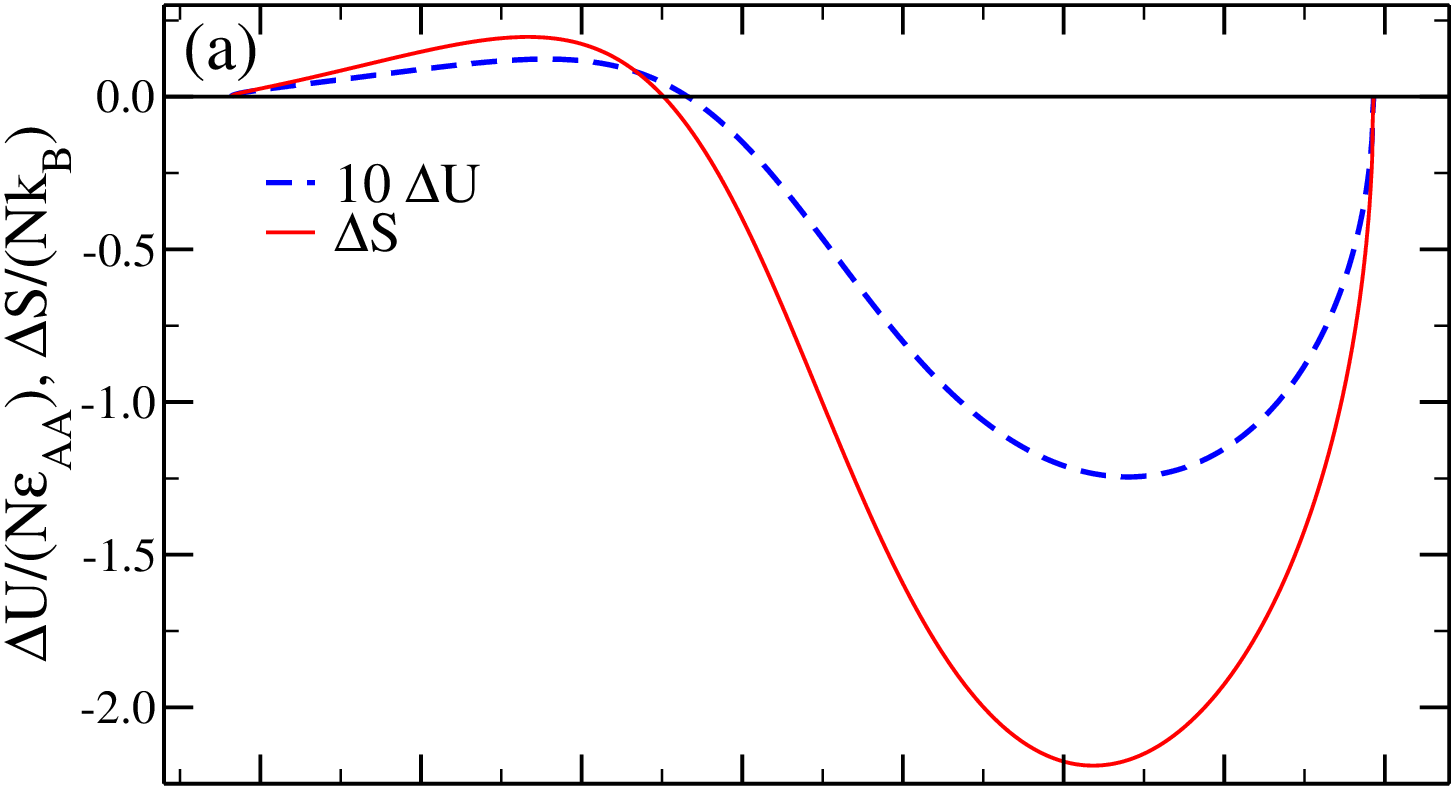}}
~\\[0.01cm]
\centerline{\includegraphics[width=3.3in]{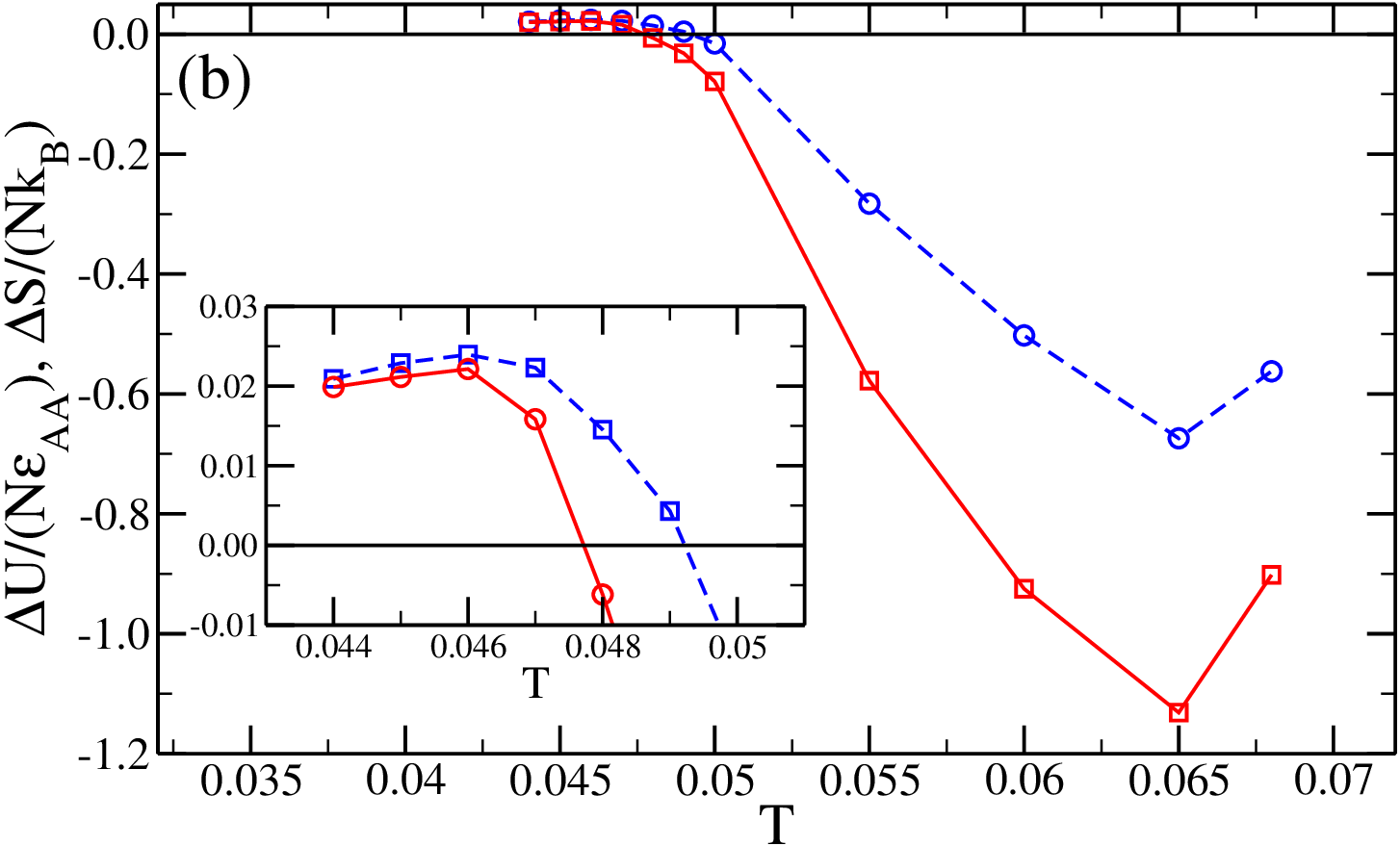}}
\caption{Difference in internal energy per particle $\Delta U /(N\epsilon_{AA})$ (dashed blue line, multiplied by 10 for clarity) 
and entropy per particle $\Delta S/(NK_B)$ (solid red line) between the liquid and the vapor phases. (a) Results from Wertheim theory. (b) Simulation results. Inset: Blow-up of the region near the lower critical point where the vapor has lower energy and lower entropy than the liquid.}
\label{fig:diffus}
\end{figure}

The theoretical evaluation of the differences in entropy and in internal energy between the liquid and the vapor phases, Fig.~\ref{fig:diffus}, is illuminating. It clarifies the different nature of the two critical points.  Just below the upper critical point the liquid has, as usual,  lower entropy and lower energy than the vapor. On the other hand, close to the lower critical point, it is the vapor that possesses lower energy and lower entropy. The self-assembled  ring clusters are very stable energetically and more ordered
conformationally. The loss of entropy associated with closing a chain into a ring is compensated by 
the  energetic stabilization introduced by the additional bond. 
This  facilitates the replacement of chains by rings in the vapor phase at low $T$. Rings have much lower energy and entropy than chains, causing the inversion of the usual order relation between the values of these quantities in coexisting phases. 

{\it Simulation results:} The phase equilibria is investigated with Successive Umbrella Sampling (SUS) simulations~\cite{sus} in the grand canonical ensemble with specific moves significantly speeding up equilibration, allowing for the  evaluation of the density of states $P(\rho, U)$,  at fixed activity $z$,  $T$ and volume $V$. Specifically, we have implemented the aggregation-volume bias~\cite{avbmc} (AVB) algorithm, its
specialization to the case of chain forming patchy colloids  (end-hopping move~\cite{lisbona_lungo}) 
and  a novel cluster-swap move which attempts to swap chains and rings of the same size.
All these methods are discussed in details in the S.I.  With all these techniques and 
a significant amount of computation (of the order of 1000  months on a single core)
we have been able to study the behavior of the system down to $T=0.039$. 
The box side is $L=14$ for $T > 0.055$ and $L=22.2$ otherwise. We use the standard Bruce-Wilding method to pinpoint the location of the critical points~\cite{bruce_wilding_energy_mixing}. In order to properly estimate the lower critical parameters we join several $P(\rho, U)$ computed at different $T$ by implementing the multiple-histogram reweighting method~\cite{multiple_histogram}. 

Panel (b) of Figs.~\ref{fig:phd}-\ref{fig:frnenj}-\ref{fig:diffus} shows the simulation results for 
the phase diagram, the  fraction of particles in rings, chain	 ends  and junctions	 and the difference in
energy and entropy along the gas-liquid coexistence curve.  In all cases, Wertheim theory qualitatively,
if not quantitatively, properly predicts the behavior of the system, confirming the unconventional theoretical
predictions. Even the  reversal of the  entropic and energetic contribution   to the transition is observed numerically.
 Beside the well known underestimate
of the coexisting liquid density characteristic of Wertheim theory, 
 the proposed extension to ring formation appear to overestimate the amount of rings
 in the sample (Fig.~\ref{fig:frnenj}). This is also confirmed by the theoretical overestimation of the amount of bonds in the system at low $T$ (see S.I.). 

{\it Conclusions:} Comparison with the simulation results shows 
that the  parameter-free Wertheim's theory for dissimilar patches, extended to include rings, 
provides correct qualitative (if not quantitative for some observables) predictions for 
 the relation between phase separation and self-assembly in complex linear structures. 
 The theory provides a powerful instrument to    control the  competition between  the formation of rings, chains and junctions and to evaluate  the resulting phase behavior.   The thermodynamic stability
 at low $T$ is shown to arise from the building up of non interacting clusters of particles with low energy and  low entropy. 
The theory also provides a theoretical foundation of the 
behavior  numerically observed  recently in models of Janus particles in three dimensions\cite{janusprl} and in 2D-simulations of limited valence particles on lattice~\cite{almarza_rings}. In both cases, the gas reentrance  is connected to the self assembly into weakly or non interacting  saturated aggregates (micelles for the Janus particles and rings for the patchy particles).

Finally we note that our results are also of indisputable value for a deep understanding of the thermodynamic of dipolar hard-spheres, the paradigmatic model of  anisotropic interactions~\cite{degennespincus}.  Despite the model's apparent simplicity, the low $T$ behavior of this system is still object of controversial interpretations.   The early predictions~\cite{2000JPCM12R411T,degennespincus}    
of a normal liquid vapor phase separation originated by an effective isotropic attractive potential, were challenged by numerical simulations~\cite{weis,tavares2d} that showed that dipoles self assemble  in chains,   branched chains  \cite{2000JPCM12R411T,campPRL,campMP} and (at low $T$) in rings~\cite{dhs_prl,dhs_soft_matter,susceptibility_prl}, the same structures reproduced in the presently investigated patchy particle model.
The similarity between the interactions and the self assembled structures formed in the DHS  and those of the patchy model under study \cite{lisbona_lungo}, can be  hopefully used  to establish a quantitative mapping between both models. The present study suggests that the absence of gas-liquid phase separation in DHS, despite the branching, could be a consequence of extensive ring formation. 

{\it Acknowledgments:} J.M.T. acknowledges financial support from the Portuguese Foundation for Science and Technology (FCT) under Contract Nos. PEstOE/FIS/UI0618/2011 and 
PTDC/FIS/098254/2008. L.R. and F.S. acknowledge support from ERC-PATCHYCOLLOIDS and
MIUR-PRIN. We thank J. Russo and F. Romano for fruitful discussions.

\end{document}